\documentclass[preprint,aps,pra,showpacs]{revtex4}
\usepackage{tabularx}
\usepackage{dcolumn}
\usepackage{graphics}% Include figure files
\usepackage{bm}% bold matha
\usepackage[dvips]{graphicx}
\usepackage[dvips]{rotating}
\usepackage[latin1]{inputenc}
\usepackage{dcolumn}
\usepackage{tabularx}
\usepackage{amsmath}
\usepackage{amssymb}
\begin{document}
\draft

\title{Escape of quantum particles from an open cavity}

%\author{H. J. Zhao and  M. L. Du} \email{
%duml@itp.ac.cn}
%\address{Institute of Theoretical Physics, Chinese Academy of Sciences,
%P.O.Box 2735, Beijing 100080, China}
\author{H. J. Zhao$^1$ and  M. L. Du$^2$ }\email{
duml@itp.ac.cn} \affiliation{$^1$School of Physics and Information
Science, Shanxi Normal University, Linfen 041004, China}
\affiliation{$^2$Institute of Theoretical Physics, Chinese Academy
of Sciences, P. O. Box 2735, Beijing 100080, China}

\date{\today}

\begin{abstract}
A negative ion irradiated by a laser provides a coherent source of electrons propagating out from the location of the negative ion. We study the total escape rates of the electrons when the negative ion is placed inside an open cavity in the shape of a wedge. We show the wedge induces significant oscillations in the total escape rates because of quantum interference effects. In particular, we show, for a wedge with an opening angle of $\pi/N$, where $N$ is an arbitrary positive integer, there are $(2N-1)$ induced oscillations in the rates. As a demonstration, the case for a wedge with an opening angle $\pi/5$ is calculated and analyzed in detail.
\par


\par
%\keywords {Escape rate; H$\mathrm{\acute{e}}$non-Heiles}

\end{abstract}

 \maketitle

%\begin{multicols}{2}

\section{Introduction}

Recently the escape of classical particles from a vase-shaped cavity was studied by Hansen, Mitchell and Delos\cite{Hansen}. They demonstrated how the fractal structure can significantly affects the escape. The system is simple but it emulates other more complex systems such as the ionization of hydrogen in parallel electric and magnetic fields\cite{Mitchell1,Mitchell2}.
A negative ion irradiated by a laser provides a coherent source of electron at the position of the negation ion\cite{Bryant}. When the negative ion is placed inside an open cavity, the quantum electron will eventually escape from the cavity. In this paper, we study the photodetachment process of H$^-$ inside a wedge and the subsequent escape of the quantum electrons. In particular, we are interesting in calculating \emph{total escape rates}. Using a procedure similar to the previous case\cite{Du89}, it is straightforward to show that the total escape rates in the present case are equivalent to the total photodetachment cross sections in the wedge. Henceforth, we will speak of cross sections instead of total escape rates.
Closely related to the present problem, we note Yang \emph{et al} \cite{Yang} recently studied the photodetachment of H$^-$ near a reflecting surface. Subsequently Afaq \emph{et al} \cite{Afaq} developed a theoretical imaging
method and applied it to study the same system.

The plan of the article is as follows. In Sec. II we describe the formulas for the photodetachment cross sections inside a wedge. The formulas are valid for any linear laser polarization relative to the  wedge with an opening angle of $\pi/N$, where $N$ is an arbitrary positive integer. In Sec.III, as an example, we calculate and analyze the photo-detachment cross sections inside a wedge with an opening angle of $\pi/5$. The dependence of the cross sections on the laser polarization is studied in detail. Conclusions and perspective are given in Sec.IV.

%%%%%%%%%%%%%%%%%%%%%%%%%%%%%%%%%%%%%%%%%%%%%%%%%%%%%%%%%%%%%%%%%%%%%%%%%%%
\section{formulas for photodetachment cross sections}

In Fig.1 we show a cross-sectional plane of the system.
The negative ion is assumed to be at the origin.
$\alpha$ is the opening angle of the wedge. $\rho$ is the distance between the
wedge axis and the negative ion, and $\beta$ is the declination angle
relative to the left surface of the wedge. It is convenient to choose the coordinate system such that the $z$-axis is parallel to the wedge axis. The $x$-axis is perpendicular to left surface and the $y$-axis is parallel to the left surface.
The $x-y$ plane is on the cross-sectional plane.

According to the physical picture of closed-orbit theory (COT)\cite{Du4}, when the active electron is detached by a laser, the active electron and the associated wave propagates out from the negative ion in all directions. The electron trajectories and wave fronts follow straight lines inside the wedge until they are reflected by the surfaces of the wedge. The electron may return to the region of the negative ion after several reflections. If so, the returning electron wave will interfere with the initial outgoing electron wave and induce oscillations in the total photodetachment cross sections. Closed-orbit theory relates the cross sections to all the closed-orbits of detached-electron and provides a recipe to calculate the cross sections based on the closed-orbits and their associated properties.

Because the trajectories are reflected by the surfaces of the wedge, it is clear that all the closed-orbits which go out from the position of the negative ion and later return to the position of negative ion must be on the $x-y$ plane.
To find closed-orbits in a wedge with an arbitrary opening angle, we can launch a large number of trajectories going out from the origin on the $x-y$ plane and keep track of the trajectories as they propagate and get reflected inside the wedge. If a group of nearby trajectories comes back to the region of the negative ion, the search can be refined to find the closed-orbit. This procedure has been previously used to find the closed-orbits for an atom in a magnetic field\cite{Du4}. In Fig.1 we show a group of trajectories near a closed-orbit (solid line) returning to the region of the negative ion.

According to COT\cite{Du4}, the total photodetachment cross section can be written as
\begin{equation}\label{1}
  \sigma(E)=\sigma_0(E)+\sigma_{osc}(E),
\end{equation}
where $\sigma_0(E)=16\sqrt{2}B^2\pi^2E^{3/2}/3c(E_b+E)^3$  represents the
photodetachment cross section of H$^-$ without the wedge,
$B=0.31522$ is related to the normalization of the initial bound
state $\Psi_i$ of $\textrm{H}^-$, $c$ is the speed of light and its
value is approximately 137 a.u.. $\sigma_{osc}(E)$ represents the
oscillating part of the cross section associated with various detached-electron
closed-orbits which will be detailed below. Closed-orbit theory gives
\begin{equation}\label{2}
\sigma_{osc}(E)=-\frac{4(E+E_b)\pi}{c}{\rm Im}\langle
D\Psi_i|\Psi_{ret}\rangle,
\end{equation}
where $\Psi_i$ is the initial bound state wave function of
$\textrm{H}^-$ and is given by $\Psi_i=B\frac{e^{-k_b r}}{r}$ in the
present one active electron approximation for photodetachment,
$k_b=\sqrt{2E_b}$, $E_b$ is the binding energy and it is
approximately $0.754eV$, and $D$ is a dipole operator which will be specified  below. We will treat the general linear polarization case.
If $\hat{\bf \epsilon}$ is the direction of the laser polarization and $(r,\theta,\phi)$ is the spherical coordinate of the detached-electron, the dipole operator can be written as
$D=r\hat{\bf r} \cdot \hat{\bf \epsilon}$. If we use $(\theta_L,\phi_L)$ to denote the spherical angles of the laser polarization direction, then $\hat{\bf r} \cdot \hat{\bf \epsilon}=f(\theta,\phi;\theta_L,\phi_L)$ and\cite{Du2006}
\begin{equation}\label{3}
   f(\theta,\phi;\theta_L,\phi_L) = \cos(\theta)\cos(\theta_L)+\sin(\theta)\sin(\theta_L)\cos(\phi-\phi_L).
\end{equation}

The returning wave $\Psi_{ret}$ near the negative ion in Eq.(2) is represented as a sum over closed-orbits and is related to the initial outgoing detached-electron wave $\Psi_{out}(r,\theta,\phi)$ as
\begin{equation}\label{4}
    \Psi_{ret}(\mathbf{r})=\sum_{j}
    \Psi_{out}A_{j}e^{i(S_{j}-\mu_{j}\pi/2)},
    %\Psi_{out}A_{j}e^{i(S_{j}-\mu_{j}\pi/2)},
\end{equation}
where the sum runs over all the electron closed-orbits going out
from and returning to the nucleus, $S_{j}$,$A_{j}$ and $\mu_{j}$
are, respectively, the action, amplitude and Maslov index of the
closed-orbit $j$;
$\Psi_{out}(r,\theta,\phi)=-\frac{4Bk^2i}{(k_b^2+k^2)^2}h_1^{(1)}(kr)f(\theta,\phi;\theta_L,\phi_L)$
is the initial outgoing electron wave from the negative
ion\cite{Du2006}, where $k=\sqrt{2E}$, and $E$ is the energy of
electron after detachment.

To calculate the returning wave function associated with the closed
orbits, we draw a sphere of radius $R$. $R$ is large enough so that the
asymptotic approximation $h_1^{(1)}(kr)\approx
-e^{ikr}/kr$ is valid. The sphere must also be
small enough so that the distance between the surface of the sphere and surfaces of   the wedge is large. The direct outgoing wave for the
detached-electron on the surface of the sphere is then
\begin{equation}\label{5}
    \Psi_{out}(R, \theta, \phi)=\frac{4Bk^2i}{(k_b^2+k^2)^2}f(\theta,\phi;\theta_L,\phi_L)\frac{e^{ikR}}{kR}.
\end{equation}
As the wave in Eq.(5) propagates out from the sphere along a
closed-orbit as illustrated in Fig.1, its amplitude and phase vary. We use $A_je^{i(S_j-\mu_j\pi/2)}$ to count for these variations in the wave function when it comes back via closed-orbit $j$. $A_j$ is a measure of the
divergence of adjacent trajectories from the closed-orbit $j$.
The expression in spherical coordinate is most convenient to calculate the amplitude $A_{j}$ in the present case. A simple manipulation gives
\begin{equation}\label{6}
    A_j=\frac{R}{R+T_jk}=\frac{R}{R+L_j},
\end{equation}
where $L_j=kT_j$ is the length of the $j$th trajectory, and $S_j$ is the classical action along the $j$th closed-orbit which is defined
\begin{equation}\label{7}
    S_j=\oint \mathbf{k}\cdot d\mathbf{l}.
\end{equation}
For the present system, we have $S_j=kL_j$.

In additional to the accumulated phase change of detached-electron wave which is expressed in the action, there is also a sudden phase change in the wave function accompanying each reflection by the surfaces of the wedge. We denote the phase loss of the wave function accompanying each reflection by $\Delta$. In our numerical calculations throughout this article, we set $\Delta$ to $\pi$ corresponding to the "hard" wedge wall case\cite{Afaq}. If the number of reflections of $j$th closed-orbit is $m_j$, then $\mu_{j}=2m_j$ in the "hard" wall case and
the $j$th returning wave inside the
small sphere is given by
\begin{equation}\label{8}
\Psi^j_{ret}({\bf r})
=\frac{4iBkf(\theta^j_{out},\phi^j_{out};\theta_L,\phi_L)}
{(k_b^2+k^2)^2}
      \frac{e^{i(kL_j- m_j\Delta)}}{L_j}e^{i\mathbf{k}^j_{ret}\cdot\mathbf{r}},
\end{equation}
where $\hbar\mathbf{k}^j_{ret}$ is the momentum of the returning
electron near the atom and $(\theta^j_{out},\phi^j_{out})$ are
the spherical coordinates defining the outgoing direction of the $j$th
closed-orbit. The returning wave function inside
the sphere can be approximated by a plane wave which can be written
as
\begin{equation}\label{9}
      \Psi^j_{ret}({\bf r})=N_je^{i\mathbf{k}^j_{ret}\cdot\mathbf{r}},
\end{equation}
where $N_j$ can be obtained from Eq.(8) and Eq.(9) as
\begin{equation}\label{10}
N_j=\frac{4iBkf(\theta^j_{out},\phi^j_{out};\theta_L,\phi_L)}
{(k_b^2+k^2)^2}
      \frac{e^{i(kL_j- m_j\Delta)}}{L_j}
%      N_j=\frac{4iBkf(\theta^j_{out},\phi^j_{out};\theta_L,\phi_L)}{(k_b^2+k^2)^2}\frac{e^{i(kL_j-m_j\pi)}{L_j},
\end{equation}

With the help of the integral in Appendix A, we immediately get the result for the following overlap integral
\begin{equation}\label{11}
\langle D\Psi_i|\Psi^j_{ret}({\bf r}) \rangle
=-\frac{32\pi k^2B^2f(\theta^j_{out},\phi^j_{out};\theta_L,\phi_L)\hat{\bf \epsilon}\cdot\hat{\bf k}^j_{ret}}{(k_b^2+k^2)^4L_j}e^{i(kL_j-m_j\Delta)}.
\end{equation}

Therefore the total photodetachment cross section can be written compactly for linear polarization in the direction of $\hat{\bf \epsilon}$ as
\begin{eqnarray}\label{12}
 \sigma(E)=\sigma_0(E)+\frac{16\pi^2B^2E}{c(E_b+E)^3} \sum_j\frac{1}{L_j} (\hat{\bf \epsilon}\cdot\hat{\bf k}^j_{out}) (\hat{\bf \epsilon}\cdot\hat{\bf k}^j_{ret} ) \sin(kL_j-m_j\Delta),
\end{eqnarray}
where the summation is over all closed-orbits going out from and returning to the position of the negative ion.
If we use $(\theta^j_{ret},\phi^j_{ret})$ to denote the spherical angles of the returning momentum vector $\hbar{\bf k}^j_{ret}$, we have
$(\hat{\bf \epsilon}\cdot\hat{\bf k}^j_{out}) (\hat{\bf \epsilon}\cdot\hat{\bf k}^j_{ret} )=f(\theta^j_{out},\phi^j_{out};\theta_L,\phi_L)
f(\theta^j_{ret},\phi^j_{ret};\theta_L,\phi_L)$.
The total photodetachment cross section in a wedge can be written alternatively as
\begin{eqnarray}\label{13}
\sigma(E)=\sigma_0(E)+\frac{16\pi^2B^2E}{c(E_b+E)^3} \sum_j\frac{f(\theta^j_{out},\phi^j_{out};\theta_L,\phi_L)
f(\theta^j_{ret},\phi^j_{ret};\theta_L,\phi_L)}{L_j}\sin(kL_j-m_j\Delta),
\end{eqnarray}
where the same summation over all the closed-orbits is implied.
While numerical search may be necessary to find all the closed-orbits for a wedge with an arbitrary opening angle, the problem of finding all the closed-orbits is simplified for a wedge with an opening angle $\alpha=\pi/N$, where $N$ is an arbitrary positive integer. Indeed, for a wedge with an opening angle $\alpha=\pi/N$, we find there are $(2N-1)$ closed-orbits and these closed-orbits can be found by using a method of images as demonstrated in a recent study on spontaneous emission rate of an atom in a metallic wedge\cite{Zhao2}. Here we give only a brief summary of the results.

For a wedge with an opening angle of $\pi/N$, there are $(2N-1)$ closed-orbits. All the closed-orbits have polar angles $\theta_{out}=\theta_{ret}=\pi/2$. Their outgoing azimuthal angles are given by
\begin{equation}\label{14}
     \phi^j_{out}=\left \{
    \begin{array}{cc}
    \frac{(j+1)\pi}{2N}, &(j=2n-1;~~n=1,2,\cdots,N);\\
    \frac{j\pi}{2N}+\beta,
    &(j=2n;~~n=1,2,\cdots,N-1).
    \end{array}
    \right.
\end{equation}
The corresponding azimuthal angles of the returning momentum $\hbar\mathbf{k}^j_{ret}$ are given by
\begin{equation}\label{15}
     \phi^j_{ret}=\left \{
    \begin{array}{cc}
    \ \phi^j_{out}+\pi, &(j=2n-1;~~n=1,2,\cdots,N);\\  \phi^{2N-j}_{out}+\pi,
    &(j=2n;~~n=1,2,\cdots,N-1).
    \end{array}
    \right.
\end{equation}

The lengths of the closed-orbits are
\begin{equation}\label{16}
    L_j=2\rho|\sin(\phi^j_{out}-\beta)|,~~~j=1,...,2N-1.
\end{equation}
The reflection numbers are
\begin{equation}\label{17}
     m_j=\left \{
    \begin{array}{cc}
    \ j, &~(j=1,2,\cdots,N);\\  2N-j,
    &~(j=N+1, N+2,\cdots,2N-1).
    \end{array}
    \right.
\end{equation}

When the above properties of the closed-orbits are used in Eq.(13), the total cross section for the $x$-polarization can be worked as
\begin{eqnarray}\label{18}
  \nonumber
     \sigma^x(E)&=&\sigma_0(E)+\frac{3\sigma_{0}}{2k\rho\sin\beta}\sin(2k\rho\sin\beta)
       +\sum_{n=1}^{N-1}\frac{3\sigma_{0}\cos^2(\frac{n\pi}{N})}{2k\rho\sin(\frac{n\pi}{N}-\beta)}\sin[2k\rho\sin(\frac{n\pi}{N}-\beta)]\\
     &~&+\sum_{n=1}^{N-1}\frac{3\sigma_{0}\cos(\frac{n\pi}{N}+\beta)\cos(\frac{n\pi}{N}-\beta)}{2k\rho\sin(\frac{n\pi}{N})}
    \sin[2k\rho\sin(\frac{n\pi}{N})].
\end{eqnarray}
The total cross section for the $y$-polarization is
\begin{eqnarray}\label{19}
      \nonumber \sigma^y(E)&=&\sigma_0(E)+\sum_{n=1}^{N-1}\frac{3\sigma_{0}\sin^2(\frac{n\pi}{N})}{2k\rho\sin(\frac{n\pi}{N}-\beta)}
\sin[2k\rho\sin(\frac{n\pi}{N}-\beta)]\\
    &~&-\sum_{n=1}^{N-1}\frac{3\sigma_{0}\sin(\frac{n\pi}{N}+\beta)\sin(\frac{n\pi}{N}-\beta)}{2k\rho\sin(\frac{n\pi}{N})}
     \sin[2k\rho\sin(\frac{n\pi}{N})].
\end{eqnarray}
The oscillatory contributions from the closed-orbits vanishes for the $z$-polarization. Therefore the total cross section for the $z$-polarization is the same as the cross section without the wedge,
\begin{equation}\label{20}
    \sigma^z(E)=\sigma_0(E).
\end{equation}

Because there are $(2N-1)$ closed-orbits in a wedge with an opening angle of $\pi/N$, there are usually $(2N-2)$ oscillatory terms in the cross sections. In the above formulas we observe $(2N-1)$ oscillations in the total cross section for the $x$-polarization case. For the $y$-polarization, the contribution from the closed-orbit perpendicular to the $y$-axis vanishes and there are $(2N-2)$ oscillations in the cross section. For the $z$-polarization, all the $(2N-1)$ closed-orbits are perpendicular to the laser polarization, all the $(2N-1)$ oscillatory terms vanish and consequently there is no oscillation.

\section{analysis of cross sections for a $\pi/5$ wedge}

As an example, we now discuss the photodetachment cross sections inside a wedge with an opening angle $\pi/5$.
We arbitrarily set $\beta=\pi/15$ for the purposes of numerical calculations. For this wedge, there are nine closed-orbits leaving from and returning to the position of the negative ion. All the closed-orbits are shown in Fig.2. They are labeled by
$j$ in the ascending order of outgoing azimuthal angle $\phi^j_{out}$. Because all the closed-orbits must be on the cross-sectional plane, we have $\theta^j_{out}=\theta^j_{ret}=\pi/2,j=1,...,9$. The first closed-orbit leaves the atom in a direction with an azimuthal angle of $\phi^1_{out}=\pi/5$ relative to the $x$-axis, and it returns to the atom after being reflected perpendicularly by the right wedge surface; the second closed-orbit leaves the atom in a direction with a larger  azimuthal angle than the first closed-orbit, it is first reflected by the right wedge surface and then by the left wedge surface before it returns to the atom; the third closed-orbit is first reflected by
the right wedge surface, it then travels perpendicularly to the left wedge surface and retraces back to the atom. Similar descriptions can be given to the other closed-orbits. We also find closed-orbit two (four) and eight (six) have the same path but their propagation directions are opposite.

For each of the nine closed-orbits, we have evaluated the outgoing azimuthal angle $\phi^j_{out}$, the azimuthal angle of the returning momentum $\phi^j_{ret}$, the reflection number $m_j$ and the length $L_j$. The properties are summarized in Table I.

\begin{table}\label{I}
\caption{Properties of the closed-orbits inside the wedge $\alpha=\pi/5$ with
$\beta=\pi/15$.  }
\begin{tabular}{c c c c c c}
\hline\hline
  Label & $~~\phi_{out}^j~~$ & $~~\phi_{ret}^j~~$ & $~~m_j~~$ & $~~L_j~~$ \\
  \hline

  1 & $\frac{\pi}{5}$ & $\frac{6\pi}{5}$  & 1 & $2\rho|\sin(\frac{2\pi}{15})|$ \\

  2 & $\frac{4\pi}{15}$ & $\frac{28\pi}{15}$  & 2 & $2\rho|\sin(\frac{\pi}{5})|$ \\

  3 & $\frac{2\pi}{5}$ & $\frac{7\pi}{5}$  & 3 & $2\rho|\sin(\frac{\pi}{3})|$ \\

  4 & $\frac{7\pi}{15}$ & $\frac{5\pi}{3}$  & 4& $2\rho|\sin(\frac{2\pi}{5})|$\\

  5 & $\frac{3\pi}{5}$ & $\frac{8\pi}{5}$  & 5 & $2\rho|\sin(\frac{8\pi}{15})|$\\

  6 & $\frac{2\pi}{3}$ & $\frac{22\pi}{15}$  & 4 & $2\rho|\sin(\frac{3\pi}{5})|$ \\

 7 & $\frac{4\pi}{5}$ & $\frac{9\pi}{5}$  & 3 & $2\rho|\sin(\frac{11\pi}{15})|$ \\

 8 & $\frac{13\pi}{15}$ & $\frac{19\pi}{15}$  & 2 & $2\rho|\sin(\frac{4\pi}{5})|$ \\
 9 & $\pi$ & $0$  & 1 & $2\rho|\sin(\frac{14\pi}{15})|$\\
  \hline\hline
\end{tabular}
\end{table}

In Fig.3 we show the total photodetachment cross sections for the
$x$-polarization and for the $y$-polarization
as a function of photon energy
$E_{photon}=E+E_b$ (solid lines) for the wedge $\alpha=\pi/5$ with
the negative ion fixed by $\beta=\pi/15$ and $\rho=200a_0$, where $a_0$ is the
Bohr radius. Compared to the smooth cross sections (dashed lines)
in the absence of the wedge, the photodetachment
cross sections inside the wedge are quite oscillatory. Furthermore, even the closed-orbits are the same in both the $x$-polarization and $y$-polarization cases, we observe that the oscillation patterns in the cross sections are very different. For example, in Fig.3(a) for the
$x$-polarization case, the oscillation pattern appears to be dominated by low frequency oscillations while in Fig.3(b) the oscillation pattern in the $y$-polarization case appears to be dominated by higher frequency oscillations. This suggests a strong dependence of the photodetachment cross section on the laser polarization direction. To explain these features, we display in Fig.4 the oscillatory part of
the cross sections and the contribution from each closed-orbit in both polarization cases. From Fig.4(a) it is clear that in the $x$-polarization case the shorter closed-orbits such as 1,2,7,8,9 make larger contributions and the longer closed-orbits such as 3,4,5,6 make smaller contributions. Shorter closed-orbits are associated with low frequency oscillations and longer closed-orbits are associated with higher frequency oscillations. When these contributions are added together, the overall oscillation pattern appears to be dominated by low frequency oscillations for $x$-polarization as in Fig.3(a). In Fig.4(b) for the $y$-polarization we observe that all the closed-orbits except 9 are equally important. This leads to a more oscillatory pattern in the cross section in Fig.3(b).

Now we consider the dependence of the photodetachment cross section on  the negative ion position. In Figs. 5(a) and 5(c) we fixed the angle $\beta=\pi/15$ and calculated the cross sections as functions of the  radial distance of the negative ion from the wedge axis in both the $x$-polarization and $y$-polarization cases. In both cases,we observe the cross sections display oscillations with damping oscillation amplitudes as $\rho$ becomes larger. In Fig.5(b) and Fig.5(d) we fixed $\rho=200a_0$ and show the cross sections as the angle $\beta$ is varied in both the $x$-polarization and $y$-polarization cases. It is noted that for the $x$ polarization, the oscillation amplitude is enhanced when the negative ion approaches either the left wedge surface or the right wedge surface.
But in the $y$-polarization case, the oscillation amplitude is enhanced when the negative ion moves close to the right wedge surface but is reduced when the negative ion is close to the left wedge surface.

Since the cross section is influenced strongly by the laser polarization,  in Fig.6 we show the oscillating part of the cross section
as a function of the laser polarization
direction. The photon energy was set to 1.0eV, the radial distance $\rho$ was set to $200a_0$ and the angle $\beta$ was set to $\pi/15$. We can conclude from this figure that the oscillating part or the quantum interference is largest when the polar angle of the laser polarization is on the cross-sectional plane on which all the closed-orbits live or  $\theta_L=\pi/2$. When the laser polarization direction is perpendicular to the cross-sectional plane, the oscillating part of the cross section vanishes. Furthermore, one can conclude from Fig.6 that the influence on the cross section is maximized when the azimuthal angle of the laser polarization is parallel to the $x$-axis.

\section{concluding remarks }

In summary, we have studied the photodetachment process of H$^-$
inside a wedge and calculated the total escape rates of the electrons.
In the present case, the total escape rates are equal to the total photodetachment cross sections which have be calculated using COT. The cross section can be expressed as a sum of a smooth part and an oscillating part as in Eq.(1). The smooth part describes the direct escape of the electrons from the negative ion in the absence of the wedge, and the quantum interference effect of the wedge on the cross section is embodied in the oscillating part. The oscillation terms would disappear for the escape of classical particles. In particular, we find for a wedge with an opening angle $\pi/N$, where N is an arbitrary positive integer, there are $(2N-1)$ induced oscillations in the total escape rates. We have made the calculations by assuming the reflection by the wedge surfaces induces a phase loss of $\Delta=\pi$ corresponding to a "hard" wall\cite{Afaq}. Similar calculations can be carried out by assuming the reflection induces a phase loss of $\Delta=\pi/2$ corresponding to a "soft" wedge wall. The change in the value of $\Delta$ will lead to some changes in the details of the escape rates but should not change the main conclusions.

\begin{center}
{\bf ACKNOWLEDGMENTS}
\end{center}
\vskip8pt This work was supported by NSFC grant Nos. 11074260 and 10804066, SXNSF grant No. 2009011004 and TYAL.

\begin{center}
{\bf Appendix A}
\end{center}
\setcounter{equation}{0}
\renewcommand{\theequation}{A.{\arabic{equation}}}
Here we evaluate the overlap integral needed in Eq.(11),
\begin{eqnarray}
\langle D\Psi_i({\bf r})|e^{i{\bf k}^j_{ret}\cdot {\bf r}} \rangle
    & = & \int{f(\theta,\phi;\theta_L,\phi_L)Be^{-k_b r}
          e^{i\bf{k}^j_{ret}\cdot\bf{r}}d{\bf r} }, \\
    & = & 3iB\int f(\theta,\phi;\theta_L,\phi_L)e^{-k_br} j_1(kr)
        P_1(\hat{\bf r} \cdot \hat{\bf k}^j_{ret})d{\bf r}, \\
    & = & 3iB\int_0^{\infty} e^{-k_br}j_1(kr)r^2dr
         \int f(\theta,\phi;\theta_L,\phi_L)\hat{\bf r}\cdot\hat{\bf k}^j_{ret}d\Omega, \\
    & = & 3iB\frac{2k}{(k_b^2+k^2)^2} \frac{4\pi}{3}\hat{\bf \epsilon} \cdot \hat{\bf k}^j_{ret}, \\
    & = & \frac{8iBk\pi}{(k_b^2+k^2)^2}\hat{\bf \epsilon}
           \cdot \hat{\bf k}^j_{ret}.
\end{eqnarray}
In Eq.(A.3) the integral on the radial variable was given previously\cite{Du89} and the integral on the angular variables in Eq.(A.3) can be carried out by choosing the $z$-axis in the direction of $\bf{k}^j_{ret}$,
\begin{eqnarray}
  &   & \int f(\theta,\phi;\theta_L,\phi_L)\hat{\bf r}\cdot\hat{\bf k}^j_{ret}d\Omega, \\
& = & \int [\cos(\theta)\cos(\theta_L)+\sin(\theta)\sin(\theta_L)\cos(\phi-\phi_L)]
        \cos(\theta)\sin(\theta)d\theta d\phi, \\
& = & \frac{4\pi}{3}\cos(\theta_L), \\
& = & \frac{4\pi}{3}\hat{\bf \epsilon} \cdot \hat{\bf k}^j_{ret}.
\end{eqnarray}

%\newpage

\begin{figure}
\includegraphics[scale=1.50,angle=-0]{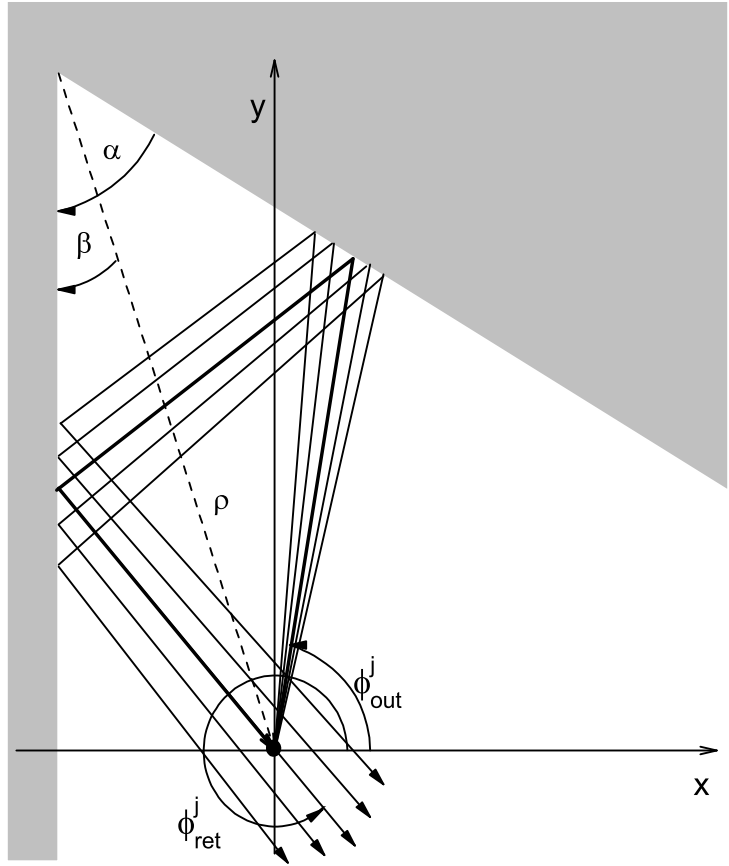}
\caption{Schematic illustration of H$^-$ photodetachment inside a wedge. The opening angle of the wedge is $\alpha$. The position of the negative ion relative to the wedge is determined by the angle $\beta$ and the distance $\rho$. We also show a group of trajectories propagating away from the H atom and finally returning to the region of the atom after being reflected twice by the wedge surfaces. The center of the group of trajectories is a closed-orbit $j$ (solid line). $\phi^j_{out}$ and $\phi^j_{ret}$ denote respectively the azimuthal angles of the outgoing and returning momenta. The quantum interference of the returning detached-electron wave associated with the closed-orbit $j$ leads to an oscillation in the cross section.}
\end{figure}

\begin{figure}
\includegraphics[scale=1.0,angle=-0]{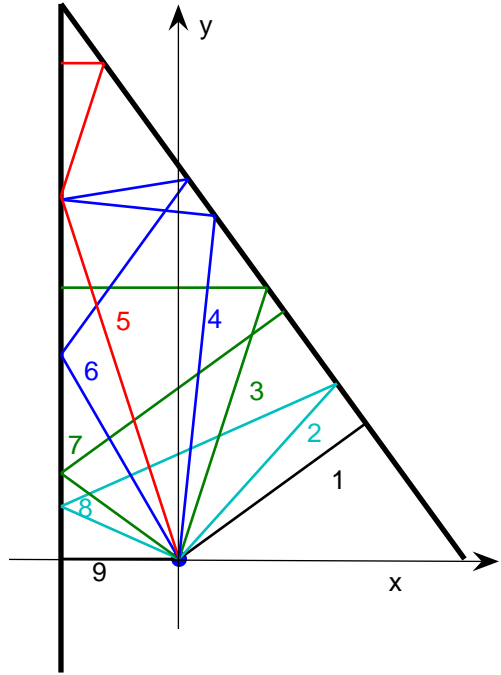}
\caption{(color online). The nine closed-orbits inside a $\pi/5$ wedge. The index numbers are written near the first segment of the closed-orbits. }
\end{figure}

\begin{figure}
\includegraphics[scale=1.5,angle=-0]{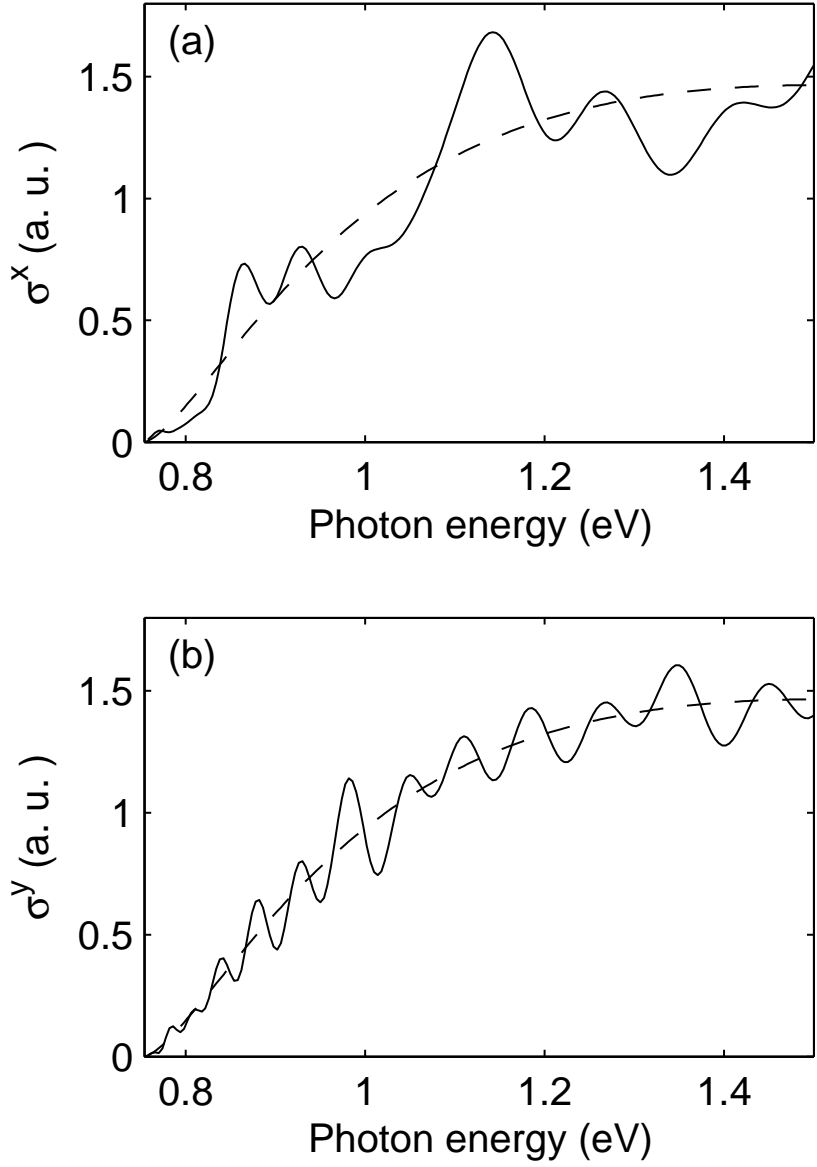}
\caption{Photodetachment cross sections (solid) for H$^-$ inside a
wedge with an opening angle $\alpha=\pi/5$. The position of the
negative ion is fixed by the angle $\beta=\pi/15$ and the distance $\rho=200a_0$, where $a_0$ is the Bohr radius. The
laser is linearly polarized in the $x$-axis direction (a)
and $y$-axis direction (b). For the purpose of comparisons, the dashed lines represent the photodetachment of a free negative ion without any wedge.}
\end{figure}

\begin{figure}
\includegraphics[scale=1.0,angle=-0]{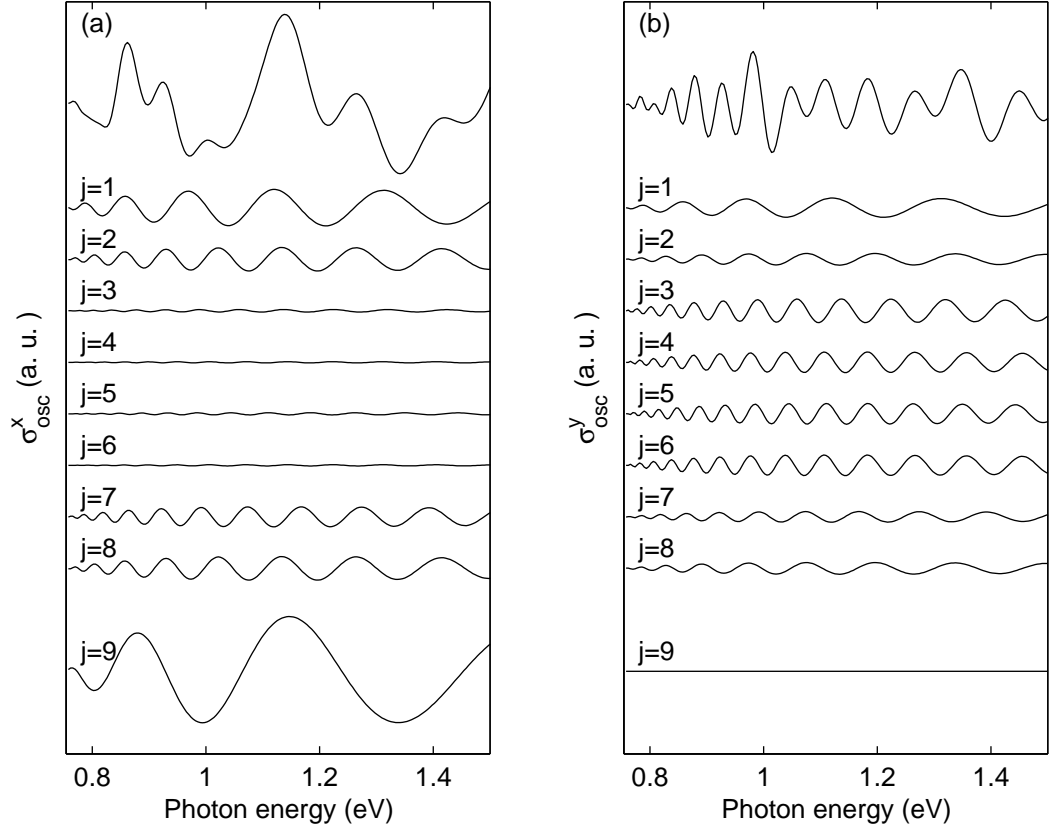}
\caption{(a) The top panel displays the total oscillating part of the cross section for $x$-polarization. The oscillatory contribution associated with each closed-orbit is displayed below from closed-orbit $j=1$ to $j=9$. (b) Similar to (a) but the polarization is in the $y$-axis direction.  }
\end{figure}

\begin{figure}
\includegraphics[scale=1.0,angle=-0]{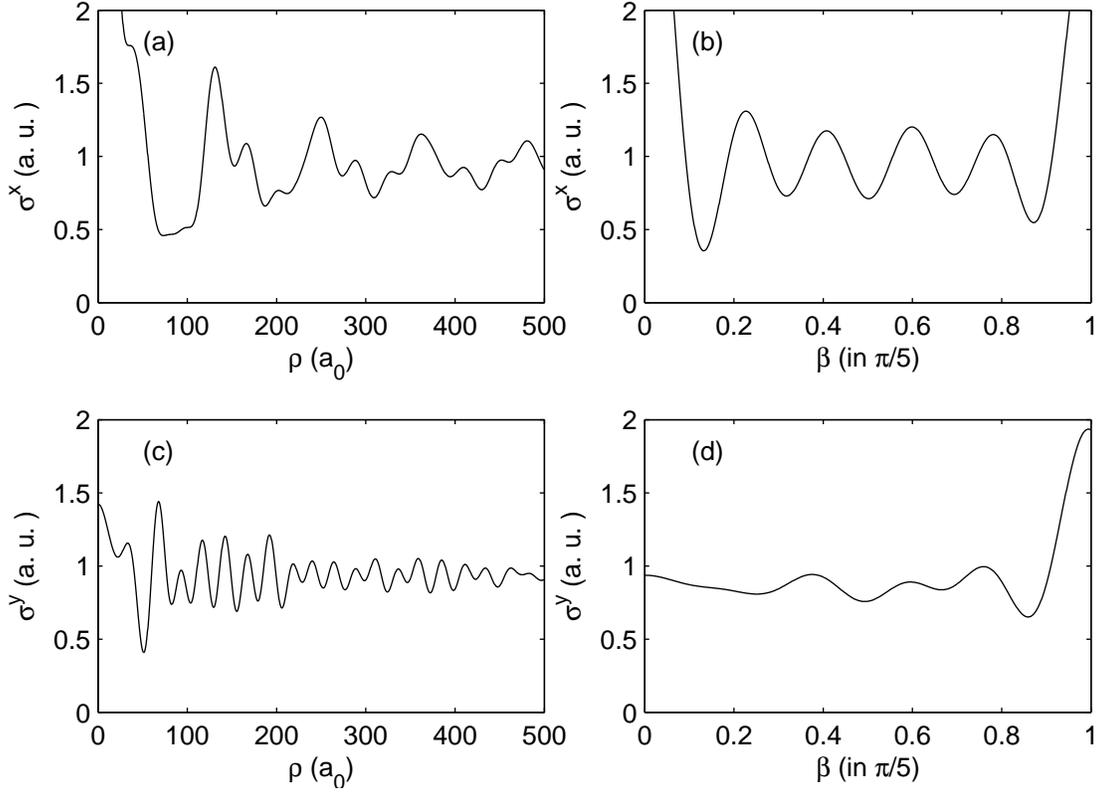}
\caption{Dependence of cross section on the position of the negative ion. (a) Variations with $\rho$ for $x$-polarization and setting $\beta=\pi/15$; (b)  variations with $\beta$ for $x$-polarization and setting $\rho=200a_0$; (c) variations with $\rho$ for $y$-polarization and setting $\beta=\pi/15$; (d) Variations with $\beta$ for $y$-polarization and setting $\rho=200a_0$.  }
\end{figure}

\begin{figure}
\includegraphics[scale=1.0,angle=-0]{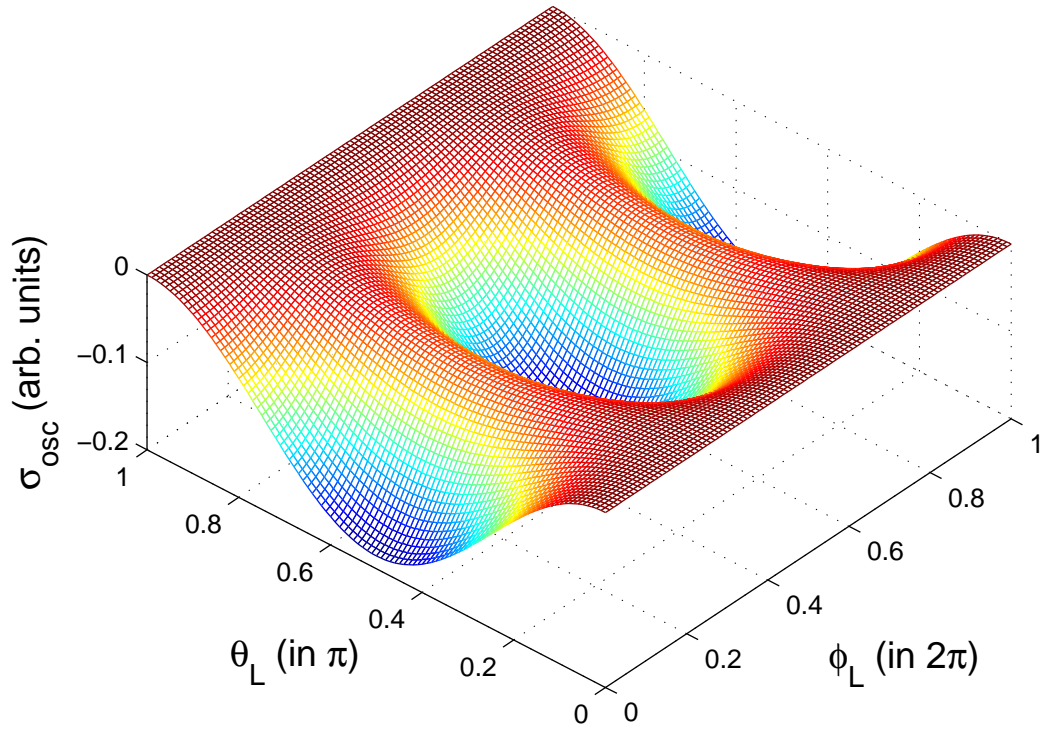}
\caption{(color online). The oscillating part of the cross section as a function of the
polarization direction. The photon energy is set to 1.0eV and the negative ion is fixed by setting $\rho=200a_0$ and $\beta=\pi/15$.
 }
\end{figure}

\end{document}